\documentclass[12pt, letterpaper]{article}
\usepackage{amsmath}
\usepackage{amsthm}
\usepackage{amssymb}
\usepackage{url}
\usepackage{slashed}
\usepackage{eucal}
\newtheorem{theorem}{Theorem}[section]
\newtheorem{rem}[theorem]{Remark}

\newcommand{\bx}{\mathbf{x}}
\newcommand{\bp}{\mathbf{p}}
\newcommand{\ba}{\mathbf{a}}

\newcommand{\BR}{\mathbb{R}}

\newcommand{\bal}{\boldsymbol{\alpha}}

\newcommand{\bma}{\left[\begin{matrix}}
\newcommand{\ema}{\end{matrix}\right]}
\newcommand{\be}{\begin{equation}}
\newcommand{\ee}{\end{equation}}
\date{}
\begin{document}
\title{Time of arrival operator in the momentum space}
\author{A. M. Schlichtinger \\
Faculty of Physics and Astronomy, University of Wroclaw\\
 pl. M. Borna 9 50-204 Wroclaw, Poland\\
 \and
A. Jadczyk \\
Laboratoire de Physique Th\'{e}orique, Universit\'{e} de Toulouse III 
and\\
Ronin Institute, Montclair, NJ 0704}
\maketitle
\begin{abstract}
It is shown that in presence of certain external fields a well defined self-adjoint time operator exists, satisfying the standard canonical commutation relations with the Hamiltonian. Examples include uniform electric and gravitational fields with nonrelativistic and relativistic Hamiltonians. The physical intepretation of these operators is proposed in terms of time of arrival in the momentum space.

\end{abstract}

\section{Introduction}

In quantum mechanics, physical quantities (observables) are represented by self-adjoint (or Hermitian) linear operators defined on a complex Hilbert space. However, it may seem puzzling that time, as one of the fundamental physical quantities, is usually considered only as a parameter entering the time evolution of states (or operators, in Heisenberg's picture). When analyzing the current state of affairs, it is difficult to resist the question: why? Physicists and philosophers again and again ask the same question: ``What is the difference between quantities that can be represented by operators, and those -- such as `amount of time' -- that cannot? And why is time the only such parametric quantity? What is special about time?'' \cite{halvorson}.

There were many attempts to answer this question, and they all refer in one or another way to the problem of time-energy uncertainty relation discussed in the classical papers  \cite{Chiny,wigner}. But the main question that needs to be answered is "Time of what"? ``Amount of time? But which time?". Certainly we have an external parameter time $t$ governing the quantum dynamics through the unitary operators $\exp(iHt/\hbar)$\footnote{For time independent Hamiltonians.}, but we may also have the internal time telling us how long it takes before some ``event" happens, for instance before some quantum observable reaches a given value.

In 1974 Kijowski \cite{Kijowski1} proposed a unique solution to the axiomatically defined problem of the time of arrival operator for a free nonrelativistic particle arriving at a flat screen (for instance a plane perpendicular to $z$-axis in the configuration space) waiting for a particle to cross it form one or the other side. Kijowski's time operator satisfies an ``almost'' canonical commutation relation with the Hamiltonian: $[H,T]=\text{sgn}(p_z)\,i\hbar.$

In his seminal paper \cite{wigner} Wigner discussed a general problem of `time of arrival at a given state', mentioning, in particular, states concentrated on a flat plane in $x$-space (while we are choosing $p$-space instead). In 1994 Mielnik \cite{mielnik94} addressed a more general "waiting screen problem". The problem is how to answer, within the formalism of quantum theory, the following question: how long it takes for a quantum particle to reach the waiting screen detector of an arbitrary shape? While for a flat screen and a free particle the answer given in \cite{Kijowski1} provides a mathematically precise and a unique solution (see, however, the clarifications in \cite{Kijowski2}), the problem with constructing a fully satisfying ``time of detection" operator remains for non-flat screens and non-free evolutions.\footnote{For a general dynamics and for any detector the problem has a simple solution within the Event Enhanced Quantum Theory that describes the dynamics of the coupling between quantum systems and classically described screens (or general detectors). But there the "time of arrival" probability distribution is not described by a linear operator. See \cite{jadczyk14} and references therein.}

{\em Yet, as we show in the present paper, in certain cases there may exist a well defined time of arrival operator, satisfying the canonical commutation relations, but {\ in the momentum\,} rather than in configuration space. Of course this would be impossible for a free particle, because for free evolutions momentum is a constant of motion.\,}

Quantum-mechanical Heisenberg's equations of motion in one dimension read (see e.g. \cite[p. 80]{hall}):
\begin{eqnarray*} \frac{dX}{dt}&=&\frac{1}{m}P(t)\\
\frac{dP}{dt}&=&-V'(X(t)).\end{eqnarray*}
The second equation, rewritten as
$$dP=-V'(X(t))\,dt$$
tells us that for nontrivial potentials the change of the momentum is proportional to the change of time. The relation becomes especially simple for static homogeneous fields, for which $V'$ is constant. For expectations values we have
$$ d\langle P(t)\rangle_\psi =-\langle V'\rangle_\psi\, dt.$$
Time can then be measured then by simply observing the change of the momentum. In this paper we are exploring this simple idea. We should mention that the case of a free particle Hamiltonian is a set of measure zero in the set of all Hamiltonians.
In our paper we demonstrate the existence of a well defined time operator on four examples: nonrelativistic Schr\"{o}dinger's particle in the uniform electric and gravitational fields, and also for a relativistic Dirac particle in such fields. In the case of a uniform gravitational field we consider general relativistic settings, where the field is described by a corresponding spacetime metric in Kottler-M\/{o}ller coordinates. In a uniform field we have a constant acceleration, and the momentum increases monotonically with time. Whether our construction can be generalized to locally non-uniform fields remains an open question. In all cases considered in this paper a simple classical interpretation of the time operators is given.

The second chapter deals with a review of the characteristics of time modeling in quantum mechanics. Difficulties, such as the Pauli theorem and the Hegerfeldt lemma (see e.g. \cite{srinivas81,halvorson} and references therein), related to the existence of the time operator in quantum theory are briefly discussed.

In the third chapter, time operators corresponding to non-relativistic Hamiltonians with interaction are presented. Their physical interpretation is also discussed.

The fourth chapter deals with the time operator in relativistic quantum theory. Time operators for a uniform electric field and a uniform gravitational field are constructed. It is shown that they meet the mathematical conditions expected from the time operator and their physical interpretation is taken into account. Their behavior under the influence of selected continuous and discrete symmetry transformations is also analyzed.

The last, fifth, chapter discusses the problem of eigenvalues of these time operators.

In the conclusions the time operator for a free massless neutrino is also mentioned.

\section{On the fundamental difficulties of creating the time operator in quantum mechanics}
In the standard formulation of quantum mechanics, an observable is a Hermitian (or self-adjoint) operator, defined on the separable complex Hilbert space $\mathcal{H},$ representing a certain physical quantity whose eigenvalues (always real) constitute the set of all obtainable results of measurement.\footnote{The measurement problem in quantum mechanics can be described by an irreversible dynamical coupling between a quantum system and a classically described measuring device, as discussed in detail in Ref. \cite{blaja95a}.} The above definition is a consequence of the historical development and of mathematical considerations (lattice theory, Gleason theorem \cite{Ame}).

In addition to self-adjointness, it is usually required that the time operator satisfies canonical commutation relation with the Hamiltonian, in analogy to conjugate pairs of position and momentum components. In view of the above requirements, numerous difficulties are encountered when trying to construct the time operator.

A typical argument against the existence of the time operator invokes ``Pauli's theorem". According to this theorem \cite[p. 3]{Gal} there cannot exist a self-adjoint time operator satisfying a commutation relation with the Hamiltonian\footnote{Throughout this article we use atomic units, in which numerical values of $\hbar$ and $c$ are $\hbar=1,\, c=1.$}
\be [\hat{T}, \hat{H}] = i \mathbb{I}, \ee
if the spectrum of the latter is half-bounded. Essentially the same result is obtained by invoking a mathematically more precise Hegerfeldt's Lemma (see \cite{srinivas81,halvorson} and references therein) or Stone - von Neumann's uniqueness theorem \cite{rosenberg}.

Difficulties in creating the time operator led to attempts to construct the corresponding object by using methods based on the positive operator-valued measure (POVM).
There are many papers (see e.g. \cite{Lud, QMM, Bush1, Bush2}), suggesting that positive operator valued measures (POVM) may be a reasonable generalization of the standard concept of observables in quantum mechanics. Moreover, the Naimark theorem \cite{Nai} shows that each symmetric operator is associated with the POVM measures induced by its generalized extensions. At the same time, the generalized concept of observables bypasses the requirement of self-adjointness, which, according to some authors (see e.g. \cite{RESPONSE}), enables the construction of a time operator in non-relativistic quantum mechanics in some very specific cases. In particular, in \cite{holevo}, A.S. Holevo discussed the method of maximal symmetric extensions to general covariant measurements including time shifts and also the photon localization problem, and in \cite{holevo2} devoted a whole section to the time observable and the ``time-energy'' uncertainty relation.

However, it should be taken into account that there are also papers devoted to criticizing the concept of POVM as observables \cite{21b, 7b}. One of the key arguments against POVM is that this construction does not solve a fundamental problem of quantum mechanics, thus failing to answer the question of what measurement is and how to model it \cite{Bell}. Therefore, POVM based time observable constructs will not be considered in this article.

Despite all these difficulties, several authors \cite{Nairobi, Chiny, oryginalAB, Recami, Tcs, bauer} have attempted to construct the time operator within non-relativistic and relativistic quantum mechanics. A special category here is represented by time superoperators acting in the Liouville space, following the Misra-Prigogine-Courbage theory of irreversibility (cf. e.g. Refs. \cite{ordonez,courbage}. Within this framework the standard Hamiltonian $H$ of the quantum system is replaced by the Liouville superoperator $L_H$ acting via the commutator on the space of density matrices. Thus if $a,b$ are eigenvalues of $H$, then $a-b$ and $b-a$ are eigenvalues of $L_H$. The spectrum of $L_H$ is therefore always symmetric with respect to the origin, and the semi-boundedness obstacle for constructing a time operator disappears. The physical and statistical interpretation of this construction is, however, in a general case, not well developed.
However, none of these papers listed provides a construction of the time operator that satisfies the conditions required from this object in this work.
\section{Nonrelativistic particle in a uniform field\label{sec:npuf}}
Restricting to just one space dimension let us consider a nonrelativistic Hamiltonian

\be H=\frac{p_x^2}{2m}+\epsilon x.\label{eq:hs}\ee
In the case of a static uniform gravitational field $\epsilon=mg$, where $g$ is acceleration due to gravity (cf. e.g. \cite{Flugge}). In the case of a uniform electric field $\epsilon=qE_x$, where $q$ is the charge value and $-E_x$ is the intensity of electric field.

One may easily see that the time operator is given by
\be \hat{t}=-\frac{p_x}{\epsilon}.\label{eq:ntg}\ee

The Hamiltonian (\ref{eq:hs}) has been studied in the discussion of Stark effect \cite{avron}, where it is shown that it is essentially selfadjoint on $C_0^\infty(\BR)$ and it has a purely absolutely continuous spectrum $\sigma(H)=(-\infty,\infty).$ In fact (see \cite{avron}) if we define the unitary operator $U$ on $L^2(\BR,dp)$ as
\be U=e^{-ip_x^3/(6m\epsilon)},\ee
then
\be UHU^*=\epsilon x,\ee
and, in fact, the pair $(T,H)$ is unitarily equivalent (using the same unitary operator $U$) to the canonical
pair ($p_x/\epsilon, \epsilon x).$

\subsection{The physical interpretation of $\hat{t}$}
In terms of the physical interpretation for a state with a defined $x$ component of momentum, such a state is the eigenstate of the time operator and its eigenvalue is the time needed to reach this value of momentum starting with a value equal to zero. For a state which is a superposition of states with a given $x$ component of momentum, the expected value of the time operator is the weighted average of such times.

\section{Time operator in relativistic quantum mechanics}
In the two cases discussed below we will use the Mandelstam-Tamm method (v. \cite{Chiny}).
The idea is as follows: let $\hat{F}$ be an observable such that $[\hat{F}, \hat{H}]^{- 1}$ exists, where $\hat{H}$ is the Hamiltonian. Then the Mandelstam-Tamm time operator is defined as:
\be\label{operator czasu Mandelstama-Tama dla granicy nierelatywist.} \hat{T}=\frac{-i\hbar}{2}\left(\hat{F}\,[\hat{F},\hat{H}]^{-1}+[\hat{F},\hat{H}]^{-1}\hat{F}\right).\ee
If $[\hat{F},\hat{H}]^{-1}$ commutes with $\hat{H}$:
\be [\hat{H},[\hat{F},\hat{H}]^{-1}]=0,\label{eq:hfh}\ee
then $\hat{T}$ automatically satisfies the relation
\be [\hat{T}, \hat{H}]=i\hbar\mathbb{I}.\label{eq:ht}\ee
\subsection{Construction of the time operator in a uniform electric field}\label{Construction of the time operator in a homogeneous electric field}

Let us consider a uniform, static electric field $\mathbf {E}$ with scalar potential
\be \phi({\bf x})=-\mathbf{E}\cdot\mathbf{x},\ee
\be E_i=-\frac{\partial \phi}{\partial x^i},\,(i=1,2,3).\ee
Thus, the Hamiltonian for the Dirac particle takes the form:
\be\label{hamiltonian elektr} \hat{H}_e=c\boldsymbol{\alpha}\cdot\hat{\mathbf{p}}+q \phi(\bx) \mathbb{I}_4+\beta m_0c^2=\hat{H}_0+q\phi(\bx) \mathbb{I}_4,\ee
Let us use the Mandelstam-Tamm method setting
\be \hat{F}=\hat{H}_0.\ee
The condition (\ref{eq:hfh}) is satisfied and the time operator satisfying the canonical commutation relations with $H$  is then given by the formula
\be\label{czas elektr} \hat{T}_e=\frac{1}{qE^2}\mathbf{E}\cdot\hat{\mathbf{p}}\mathbb{I}_4,\ee
which means that it is, up to the numerical factor, the projection of momentum on the direction of the electric field.

From the fact that the operator $\hat {T}_e$ in the momentum space is defined (and essentially self-adjoint) at the intersection of the set of integrable functions with the square of the norm, which, after being multiplied by $p_x$, retain this property, it follows that its domain is the set of functions belonging to the Schwartz space $\mathcal{S}$ \cite{reed1}.
\subsubsection{The physical interpretation of $\hat{T}_e$}
The operator $\hat{T}_e$  has a simple physical interpretation and can be derived on the basis of classical mechanics. The Lorentz force is given by
\be \mathbf{F}=\frac{d\mathbf{p}}{dt}=q(\mathbf{E}+\mathbf{v}\times\mathbf{B}).\ee
Putting $\mathbf{B}=\mathbf{0}$ one obtains
\be \frac{d\bp}{dt}=e(\mathbf{E}\cdot \bx),\ee
hence
\be \bp(t)=e\mathbf{E}t+\bp(0).\ee
Assuming $\bp(0)=\mathbf{0}$, we obtain
\be t=\frac{\bp\cdot\mathbf{E}}{E^2}.\ee
The physical interpretation of $\hat{T}$ is therefore the same as in the nonrelativistic case.
\subsection{Construction of the time operator in a uniform gravitational field}\label{Construction of the time operator in a homogeneous gravitational field based on the Hehl's Hamiltonian}

According to the principle of equivalence introduced by Einstein in the process of formulating the foundations of general relativity, the operation of a uniform gravitational field is equivalent to switching to a frame of reference moving with uniformly accelerated motion. It is enough to restrict ourselves to two spacetime dimensions. Following Ref. \cite[p. 256, eq. (140)]{Moller} let us introduce the uniformly accelerated coordinates (Kottler-M\/{o}ller coordinates) $(x,t)$ related to Minkowski coordinates $(X,T)$ via the formulas
\be X=\frac{c^2}{g}\left(\textup{cosh}\left(\frac{gt}{c}\right)-1\right)+x\textup{cosh}\left(\frac{gt}{c}\right),\ee
\be T=\frac{c}{g}\textup{sinh}\left(\frac{gt}{c}\right)+\frac{x}{c}\textup{sinh}\left(\frac{gt}{c}\right).\ee
The Minkowski's metric takes the form
\be ds^2=c^2dT^2-dX^2=c^2\left(1+\frac{gx}{c^2}\right)^2dt^2-dx^2.\ee
The classical relativistic Hamiltonian for a particle with mass $m_0$ is then (see \cite[p. 379, eq. (10)]{Moller})
\be H_g=\left(1+\frac{gx}{c^2}\right)H_0,\label{eq:m10}\ee
where
\be H_0=c(m_0^2c^2+\bp^2)^\frac12.\ee
To quantize the classical expression for the Dirac particle we can choose the simple symmetrization method. Therefore we set\footnote{In \cite{mitra} the authors consider a non-symmetrized (and thus non-selfadjoint) version of this Hamiltonian.}
\be \hat{H}_g=\frac12\left(\left(1+\frac{gx}{c^2}\right)\hat{H}_0+\hat{H}_0\left(1+\frac{gx}{c^2}\right)\right),\ee
where
\be \hat{H}_0=\ba\cdot\hat{\bp}+\beta\, m_0c^2\label{Hehl swobodny 111}.\ee
Notice that the self-adjoint boost generator $N_x$ is given by the formula (see \cite[. 37]{thaller})
\be N_x=\frac12(xH_0+H_0x).\ee
For each $\bf{b}\in \BR^3$ let $U(\bf{b})=\exp\left(\bp\cdot\bf{b}/\hbar\right)$ denote the unitary translation operator. Writing
\be \hat{H}_g=\frac{g}{c^2}\frac12\left( \left(\frac{c^2}{g}+x\right)H_0+H_0\left(\frac{c^2}{g}+x\right)\right)\ee
we find that
\be \hat{H}_g=\frac{g}{c^2}=U(c^2/g)N_xU(c^2/g)^*,\ee
and therefore $\hat{H}_g$ is also self-adjoint.

For a general $3d$ acceleration vector $\ba$ we will have
\be
\begin{split}
\hat{H}_g&=\frac12\left(\left(1+\frac{\ba\cdot\bx}{c^2}\right)\hat{H}_0+\hat{H}_0\left(1+\frac{\ba\cdot\bx}{c^2}\right)\right)\notag\\
&=\hat{H}_0+\frac{1}{2c}\left((\ba\cdot\hat{\bx})(\bal\cdot\hat{\bp})+(\bal\cdot\hat{\bp})(\ba\cdot\hat{\bx})\right)+\beta\, m_0(\ba\cdot\hat{\bx}),\label{eq:hax}
\end{split}
\ee
which agrees with the expression given in \cite{hehl} in absence of rotation.

In order to construct a corresponding time operator the Mandelstam-Tamm method will be used again. This time $\hat{F}$ will be chosen as
\be \hat{F}=(\hat{\bp}\cdot\ba)\,\mathbb{I}_4.\ee
Then the formula (\ref{operator czasu Mandelstama-Tama dla granicy nierelatywist.}) leads to
\be \hat{T}_g=-\frac{c^2}{\ba^2}\frac{(\ba\cdot\bp)(c\bal\cdot \bp+\beta m_0c^2)}{\bp^2c^2+m_0^2c^4}.\ee

\begin{rem}
In the non-relativistic limit, leaving the constant, one obtains the Hamiltonian and the time operator described earlier for the non-relativistic particle. They are expressed by the formulas
\be\label{hamiltonian Farisa} \hat{H}_\mathrm{lim}=\frac{\beta}{2m_0}\hat{\bp}^2+\beta m_0(\ba \cdot \bx),\ee
\be\label{Tnonrelat} \hat{T}_\mathrm{lim}=\frac{\beta}{m_0\|\ba\|^2}(\hat{\bp}\cdot \ba).\ee
Such operators are self-adjoint and satisfy the canonical commutation relation.
\end{rem}

However, due to the fact that in this case the operator $[\hat{H}_g, \hat{F}]^{-1}$ is not commutating with the operator $\hat{H}_g$, the desired relation of the canonical commutation of the time operator in a uniform gravitational field with the Hamiltonian is not satisfied. By a direct calculation we find that
\be [\hat{H}_g,\hat{T}_g]=i\hbar\left(1-\left(\frac{\hat{T}_g}{\tau}\right)^2\right).\label{eq:cht}\ee

The $\hat{T}_g$ is evidently self-adjoint. The domain of its self-adjointness is the entire Hilbert space, because the considered operator is bounded.  Indeed, from the definition we easily find
\be \hat{T}_g^2=\frac{c^4}{a^4}\frac{(\ba \cdot \bp )^2}{c^2p^2+m_0^2c^4},\ee
and the function on the right hand side is bounded and it approaches its supremum $c^2/a^2$ asymptotically for infinite momenta in the direction of the vector $\ba$. Therefore
\be \|\hat{T}_g\| = \frac{c}{|\ba|}.\ee
Comparing Eq. (\ref{eq:cht}) with the formula
\be \frac{d}{dx} \text{tanh}(x)=1-\text{tanh}^2(x),\ee
we find that if we define
\be \tilde{T}_g=\tau\, \text{arctanh}\left(\frac{\hat{T}_g}{\tau}\right),\ee
where
\be \tau=\frac{c}{|\ba|},\ee
then $\tilde{T}_g$, being a measurable real function of a bounded selfadjoint operator, is an unbounded self-adjoint operator \footnote{Cf. e.g. \cite[p. 79, Prop. 4.17]{schmudgen}.} with the property
\be [H_g,\tilde{T}_g]=i\hbar.\ee
\subsubsection{The physical interpretation of $\tilde{T}_g$}

One should analyze the classical Hamiltonian given by Eq. (\ref{eq:m10}).
In just one space dimension Hamilton's equations of motion are of the form
\be \frac{\partial p}{\partial t}=-\frac{\partial H}{\partial x}=-g \frac{\sqrt{c^2 m_0^2 + p^2}}{c},\ee
\be \frac{\partial x}{\partial t}=\frac{\partial H}{\partial p}=\frac{c p \left(1 + \frac{g x}{c^2}\right)}{\sqrt{c^2 m_0^2 + p^2}}.\ee
For $p\geq 0$ and such that $p(0)=0$ we get
\be p(t)=cm_0\, \textup{sinh}\left(\frac{gt}{c}\right).\ee

Classically time operator $T_g$ can be written exactly as its quantum version:
\be T_g=\frac{c^2}{g}p{\hat{H}_0^{-1}}.\ee
It is possible to express $H_0$ through $p$ as follows:
\be H_0=c \sqrt{c^2 m_0^2 \textup{cosh}\left(\frac{g t}{c^2}\right)}.\ee
Hence
\be T_g=\frac{c}{g}\textup{sech}\left(\frac{gt}{c^2}\right)\textup{sinh}\left(\frac{g t}{c}\right)=\frac{c}{g}\textup{tanh}\left(\frac{tg}{c}\right),\ee
so
\be t=\frac{c}{g} \textup{arctanh}\left(T_g \cdot\frac{g}{c}\right).\ee
The classical operator $T_g$ is the classical time $t$, in the M\/{o}ller-Kottler coordinates, required to reach the momentum $p$ starting at $p = 0$ at time $t = 0$. The physical interpretation for states other than the eigenstates of momentum can be found by decomposing them into eigenstates of momentum.

\subsection{Covariance of relativistic time operators in homogeneous external fields}

Let us write explicitly the dependence on external fields, i.e.
\be \hat{T}_e(\mathbf{E})=\frac{1}{eE^2}\mathbf{E}\cdot\hat{\mathbf{p}}\mathbb{I}_4,\ee
\be \hat{T}_g(\ba)=-\frac{c^2}{\ba^2}\frac{(\ba\cdot\hat{\bp})(c\bal\cdot \hat{\bp}+\beta m_0c^2)}{\hat{\bp}^2c^2+m_0^2c^4}.\ee
The operators $\hat{T}_e(\mathbf{E})$ and $\hat{T}_g(\ba)$ are evidently commutating with components $\hat{p}_i$ of momentum operator. In other words, these time operators are translationally invariant. It is also easy to see that they are covariant with respect to the rotation group $\mathrm{SO(3)}$. We have
\be U(R)\hat{T}_e(\mathbf{E})U(R)^*=\hat{T}_e(R\bf{E}),\ee
\be U(R)\tilde{T}_g(\mathbf{a})U(R)^*=\hat{T}_g(R\bf{a}).\ee
The parity $P$ represented by the operator $U_P$ is given by \cite[p. 105]{thaller}
\be (U_P\psi)(\bx)=\beta\psi(-\bx).\ee
Time reversal symmetry is given by the anti-unitary operator $U_T$
\be (U_T\psi)(\bx)=i\beta\alpha_2 \overline{\psi(\bx)}.\label{eq:ut}\ee
Charge conjugation symmetry is given by the anti-unitary operator $C$ defined as
\be (C\psi)(\bx)=i\beta\alpha_2\overline{\psi(\bx)}.\ee
In the case of the time operator in an external electric field we have
\be U_P\hat{T}_e(\mathbf{E})U_P^{-1}=\hat{T}_e(-\mathbf{E}),\ee
\be U_T\hat{T}_e(\mathbf{E})U_T^{-1}=\hat{T}_e(-\mathbf{E}),\ee
\be C\hat{T}_e(\mathbf{E})C^{-1}=\hat{T}_e(-\mathbf{E}).\ee
In the case of the relativistic time operator in the external gravitational field, we obtain
\be U_P\hat{T}_g(\ba)U_P^{-1}=\hat{T}_g(-\ba),\ee
\be U_T\hat{T}_g(\ba)U_T^{-1}=\hat{T}_g(-\ba),\ee
\be C\hat{T}_g(\ba)C^{-1}=\hat{T}_g(\ba).\ee

\section{Eigenvalues and eigenfunctions of time operators}
Let us recall the nonrelativistic time operators from Sec. \ref{sec:npuf}. For fields in the direction of $x$-axis they are given by
\be \hat{t}_e=\frac{p_x}{qE_x},\ee
\be \hat{t}_g=\frac{p_x}{ma_x}.\ee
Therefore their eigenfunctions coincide with eigenfunctions of the $x$-component of the momentum operator, and their eigenvalues follow immediately from the above formulas.

In the case of the relativistic Dirac particle in a uniform electric field, where
\be\hat{T}_e=\frac{1}{qE^2}\mathbf{E}\cdot\hat{\mathbf{p}}\mathbb{I}_4,\ee
again eigenfunctions of the time operator coincide with eigenfunctions of the momentum operator in the direction of the external field $\bf{E}.$

Finally, in the case of the Dirac particle in the uniform gravitational field, with
\be \hat{T}_g=-\frac{c^2}{\ba^2}\frac{(\ba\cdot\bp)(c\bal\cdot \bp+\beta m_0c^2)}{\bp^2c^2+m_0^2c^4},\ee eigenfunctions of the time operator coincide with simultaneous eigenfunctions of the momentum operator in the direction of the acceleration $\bf{a}$ and the free Hamiltonian $H_0$. The corresponding eigenvalues can then be read directly from the above formula.

\section{Conclusions}
While Kijowski's time of arrival operator provides a well defined version of the time of arrival operator for free particles, it fails to perform the desired function in presence of interactions and for non-flat screens. In this paper we have provided several examples of well defined time operators in presence of external fields acting on a ``cosmic scale'', that is non-vanishing at infinity. They have been interpreted as "time of arrival" in the momentum (rather than in configuration) space. It is an open problem if our examples can be generalized to weakly non-uniform external fields.

It is tempting to speculate that time ``flows'' only in presence of cosmic scale interactions. But there is a simple counterexample to such a conjecture - the free massless neutrino particle.
In his monograph ``The Dirac Equation"  \cite[p. 227]{thaller}, Thaller defines the operator $\hat{A}$ by the formula
\be \hat{A} = \frac{1}{2}\left(\hat{H}_0^{-1}\hat{\bp} \cdot \hat{\bx} + \hat{\bx} \cdot \hat{\bp} \hat{H}_0^{-1}\right),\ee
where
\be \hat{H}_0 = c\boldsymbol{\alpha}\cdot\hat{\bp}.\ee
The operator $\hat{A}$ is essentially self-adjoint on $C_0^{\infty}(\BR^3)^4$. At the same time $\hat{T}=-\hat{A}$ satisfies the desired commutation relation with the Hamiltonian, i.e.
\be [\hat{T}_0, \hat{H}_0]=-i\hbar\mathbb{I}_4.\ee

\end{document}